\documentclass[twocolumn,preprintnumbers,amsmath,amssymb,showkeys]{revtex4}

\usepackage[dvipdfm]{graphicx}
\usepackage{graphicx}
\usepackage{dcolumn}
\usepackage{bm}
\newcommand {\bsub} {\begin{subequations}}
\newcommand {\esub} {\end{subequations}}

\def\dgbl{\Delta g_L^b}
\def\dgbr{\Delta g_R^b}
\def\afbb{A_{FB}^b}
\def\mgkk{m_{g^{(1)}}}
\def\simgt{\,{\rlap{\lower 3.5pt\hbox{$\mathchar\sim$}}\raise 1pt\hbox{$>$}}\,}
\def\simlt{\,{\rlap{\lower 3.5pt\hbox{$\mathchar\sim$}}\raise 1pt\hbox{$<$}}\,}

\begin{document}

\preprint{OCHA-PP-300}

\title{
Kaluza-Klein gluon and 
$b$-jet forward-backward asymmetry

} 

\author{Gi-Chol Cho}
\affiliation{%
Department of Physics, Ochanomizu University,\\
Tokyo, 112-8610, Japan
}%
\author{Yoshimi Kanehata}
\affiliation{%
Graduate School of Humanities and Sciences, \\
Ochanomizu University, Tokyo, 112-8610, Japan
}%

\begin{abstract}
The forward-backward asymmetry of $b$-quark jets on the $Z$-pole 
measured at LEP/SLD experiments shows us $-2.8$-$\sigma$ deviation 
from the Standard Model (SM) prediction. 
We examine a possibility of Kaluza-Klein (KK) gluon to explain the 
$\afbb$ data in a scenario based on the warped extra dimension 
model by Randall and Sundrum. 
In this scenario, the KK gluon strongly couples to $b$-quark by an 
appropriate choice of the bulk quark mass terms. 
We find that the $\afbb$ data could be explained if the KK gluon
 mass is few hundred GeV. 
Constraints on our scenario from the hadron collider experiments are
 discussed. 

\end{abstract}

\maketitle
Standard Model (SM) of particle physics has shown a good agreement 
with the results of electroweak experiments performed on 
the $Z$-pole~\cite{:2005ema}, except for 
the forward-backward (FB) asymmetry of $b$-quark jets ($\afbb$).  
The experimental data of $\afbb$ is \cite{:2005ema}
\begin{eqnarray}
\afbb(\rm exp.) = 0.0992 \pm 0.0016, 
\label{afbexp}
\end{eqnarray}
while the SM prediction is \cite{:2005ema}
\begin{eqnarray}
\afbb(\rm SM) = 0.1037,  
\label{afbsm}
\end{eqnarray}
for the best fit of the SM. 
From (\ref{afbexp}) and (\ref{afbsm}) we find about 
$-2.8~\sigma$ deviation. 
Although it might be caused due to a lack of our understanding  
of the $b$-jet data as discussed in ref.~\cite{Amsler:2008zzb}, 
in this article we would like to examine a possibility 
of the deviation as an implication of new physics beyond the SM. 
The electroweak observables at the $Z$-pole experiments can be 
expressed in terms of the effective coupling $g_\alpha^f$ which 
denotes the interaction between $Z$ and $f_\alpha$, 
where $f$ represents 
fermion species and $\alpha(=L,R)$ is their chirality. 
The radiative corrections to $g_\alpha^f$ consist of 
the gauge boson propagator corrections (so called the oblique 
corrections) which are often parametrized by $S$ and $T$ 
\cite{Peskin:1990zt}, 
and the $Zff$ vertex correction $\Delta g_\alpha^f$.  
When the oblique correction is dominated by SM, 
the new physics contribution to the FB asymmetry, 
$\afbb({\rm NP})$,  
is given as follows~\cite{Hagiwara:1998yc}: 
\begin{eqnarray}
\afbb ({\rm NP}) 
&=& \afbb({\rm SM}) - 0.0326 \dgbl - 0.1789 \dgbr. 
~~
\label{afbb}
\end{eqnarray}
It is convenient to define the additional new physics contribution 
to $\afbb$ in the unit of $10^{-4}$ 
\begin{eqnarray}
\delta \afbb \equiv 
(\afbb({\rm NP}) - \afbb({\rm SM})) \times 10^{+4}. 
\label{delafb_exp}
\end{eqnarray}
The present experimental data (\ref{afbexp}) constrains 
the new physics contribution (\ref{delafb_exp}) as 
\begin{eqnarray}
\delta \afbb = -45 \pm 16, 
\label{delafb}
\end{eqnarray}
at the 1-$\sigma$ level. 

Several attempts have been done to explain (\ref{delafb}) 
based on various new physics models 
--  
{\it e.g.}, supersymmetry~\cite{Altarelli:2001wx}, 
extended gauge symmetry~\cite{He:2002ha}, 
extra vector-like quarks~\cite{Choudhury:2001hs}, etc. 
Contribution of Kaluza-Klein (KK) particles of the SM fields in 
a variant of warped extra dimension model by 
Randall and Sundrum (RS)~\cite{Randall:1999ee} is 
also one of the possibilities. 
In this model, the KK modes of gauge bosons 
and fermions contribute to both the oblique and $Zbb$ vertex 
corrections. 
It has been shown that the KK modes of the electroweak gauge bosons  
give significantly large contribution to the oblique parameters since 
there is no custodial symmetry in the bulk. 
As a result, the scale of KK mode $\Lambda_{\rm KK}$ is strongly
constrained from the electroweak data, say 
$\Lambda_{\rm KK}> O(10^{2-3}{\rm TeV})$, 
which leads to unwanted hierarchy between the electroweak scale  
$\Lambda_{\rm EW}\sim O(m_W)$ and 
$\Lambda_{\rm KK}$~\cite{Davoudiasl:1999tf,Davoudiasl:2000wi}.  
Such a constraint could be somewhat lowered to $O({\rm TeV})$ by 
introducing the custodial 
symmetry in the bulk, or additional contribution 
from the bulk SM fermions~\cite{Carena:2006bn,Agashe:2003zs}. 
Taking account of these constraints, the $\afbb$ puzzle has been 
studied in a variant of RS model, {\it e.g.}, in 
refs.~\cite{Djouadi:2006rk,Djouadi:2009nb}, 
where the deviation of 
$\afbb$ is explained by the mixing of the $Z$ boson and its KK states. 

In this article, we would like to study the KK gluon contribution to 
$\afbb$ in the warped extra dimension model. 
It is known that, 
in warped extra dimension model, the 4D effective coupling of 
KK gluon and fermions is determined by the overlap of their 
wavefunctions in the fifth dimension. 
With an appropriate choice of the bulk quark mass terms, 
the coupling of the KK gluon to the $b$-quark could sizably enhanced 
while the others are suppressed. 
Then, the 1-loop KK gluon exchange could shift the $Zbb$ vertex 
correction $\Delta g_\alpha^b$ without any shift to 
$\Delta g_\alpha^f(f\neq b)$. 
We find that, in this scenario, the puzzle of $\afbb$ could 
be resolved when the 1st KK gluon mass is few hundred GeV. 
As mentioned above, the KK scale is constrained to be $O({\rm TeV})$ 
taking account of the contributions of KK $W,Z$ bosons to oblique 
parameters. In this case the KK gluon mass also must be $O({\rm TeV})$
which cannot give sizable correction to $Zbb$ vertex. 
Therefore our scenario of relatively light KK gluon faces difficulty 
in models which has been known so far. 
However it is worth studying the QCD corrections to the $Zbb$ vertex 
independently from the structure of electroweak sector 
in warped extra dimension model. 

Phenomenology of the KK gluon has been studied in, {\it e.g.},  
ref.~\cite{Lillie:2007yh}, focusing on the production and decay 
at LHC. 
The KK gluon in \cite{Lillie:2007yh} is, however, assumed to 
couple strongly to the $t_R$ quark and contribution to 
$\delta \afbb$ is not considered. 

Let us briefly review the interactions of the KK gauge boson to 
fermions in the warped extra dimension model. 
The model consists of a non-factorizable geometry on $AdS_5$ 
with metric 
\begin{eqnarray}
ds^2 &=& e^{-2k|y|}\eta_{\mu\nu}dx^\mu dx^\nu - dy^2, 
\end{eqnarray}
where $y$ is the coordinate of the fifth dimension and $k$ denotes 
the $AdS_5$ curvature. 
Two 3-branes --``Planck'' and ``TeV'' branes -- 
locate at fixed points of $S^1/Z_2$ orbifold, 
$y=0$ and $y=\pi r_c$, respectively.  
The hierarchy between the Planck and Electroweak scales can be 
explained reasonably when $k r_c \approx 11$. 
In general, 
if a SM fermion $\Psi$ can propagate into the bulk,  
there is a 5D mass term $m_\Psi \bar{\Psi}\Psi$ in the 
5D action without breaking the SM gauge symmetry. 
As shown in \cite{Grossman:1999ra}, 
the 5D fermion mass $m_\Psi$ can be expressed as  
$m_\Psi = \nu_\Psi k \epsilon(y)$, where 
$\epsilon(y)$ is $+1$ for $y>0$ while $-1$ for $y<0$ to make 
the mass term to be $Z_2$-even. 
The wavefunction of the zero mode fermion, then, has the peak 
toward the Planck brane for $\nu_\Psi < -1/2$ and toward 
the TeV brane for $\nu_\Psi > -1/2$. 
The effective 4D interaction of a fermion $f^{(n)}$ and 
a gauge boson $A_\mu^{(m)}$ can be obtained by integrating 
the 5D action over $y$, 
where $f^{(n)}$ and $A_\mu^{(m)}$ are the 4D KK modes of the 5D fermion 
$\Psi$ and gauge boson $A_M$, respectively, and $n,m$ are positive 
integer. 
Then, the effective coupling of the zero mode fermion 
$f (= f^{(0)})$ and KK gauge boson $A^{(n)}$ is given as a function of 
the parameter $\nu_\Psi$. 
The generic formula of $g^{ff A^{(n)}}$ can be found, {\it e.g.}, 
in ref.~\cite{Davoudiasl:2000wi}. 
For $n=1$, the coupling $g^{ff A^{(1)}}$ can be expanded in terms  
of $\nu_\Psi$ as follows: 
\begin{eqnarray}
&&
g^{ffA^{(1)}} \approx g_{\rm SM}
\nonumber \\
&&
\times
\left\{
\begin{array}{ll}
-0.2 & (\nu_\Psi < -0.5) 
\\[2mm]
4.0 + 5.2\nu_\Psi - 4.6 \nu_\Psi^2 + 2.1\nu_\Psi^3
& (\nu_\Psi > -0.5), ~~~
\end{array}
\right. 
\label{coupling}
\end{eqnarray}
where $g_{\rm SM}$ denotes the SM gauge coupling in 4D. 
\begin{figure}[ht]
\includegraphics[width=7cm,clip]{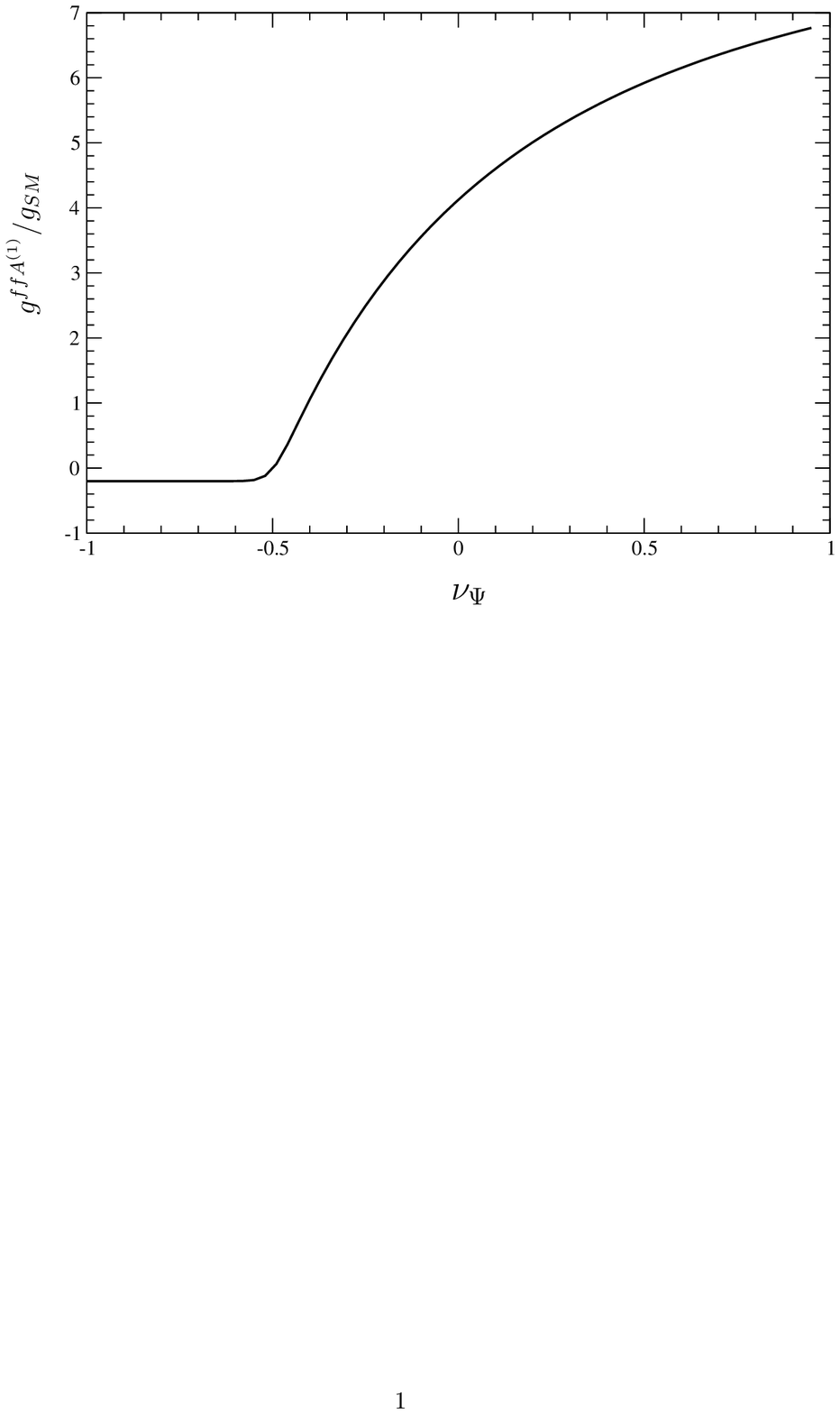}
\caption{The ratio of 4D effective coupling of 
the 1st KK mode of the gauge boson to the fermion, 
$g^{ffA^{(1)}}$, and the SM 
gauge coupling $g_{\rm SM}$ as a function of the parameter 
$\nu_\Psi$. 
}
\label{ratio_coupling}
\end{figure}
In Fig.~\ref{ratio_coupling} we depict a ratio 
$g^{ffA^{(1)}}/g_{\rm SM}$ as a function of $\nu_\Psi$. 
We find that the coupling $g^{ffA^{(1)}}$ is enhanced 
significantly for $\nu_\Psi \simgt -0.4$ 
as compared to the SM gauge coupling $g_{\rm SM}$. 
On the other hand, the coupling $g^{ffA^{(n)}}$ is highly 
suppressed for $\nu_\Psi \simlt -0.5$. 
The couplings of the higher KK mode of gauge boson with fermions 
are also highly suppressed when $\nu_\Psi \simlt -0.5$. 
In the literature, the parameter $\nu_\Psi$ is considered as 
an origin of the hierarchy of 4D Yukawa couplings. 
The values of $\nu_\Psi$ for each flavor are constrained to reproduce
the hierarchy of 4D Yukawa couplings\cite{Agashe:2003zs}. 
In our study, however, we take $\nu_\Psi$ as model parameters 
to explain the $\afbb$ data. 

Next, we examine the QCD correction to the $Zbb$ vertex 
due to the exchange of the KK gluon $g^{(n)}$ and $b^{(m)}$-quarks. 
In our study, we consider possibilities that the $b(=b^{(0)})$-quarks 
strongly couple to $g^{(n)}$, which corresponds to cases 
$\nu_{Q_{3L}}$ or $\nu_{b_R} \simgt -0.5$, where $Q_{3L}=(t_L, b_L)$. 
We set $\nu_{\rm others} \simlt -0.5$ for the other light quarks 
so that those couplings to $g^{(n)}$ are neglected. 
We do not consider the $t$-quark in the following because it does 
not contribute to $Zff$ vertex through the QCD correction. 
Then, the contributions of KK gluon to $Zbb$ vertex are determined 
by the $\nu$-parameters for $b_L,b_R$ and the KK gluon 
mass, $\mgkk$.   
From phenomenological point of view, it is useful to introduce a 
new parameter $\xi_\alpha \equiv g^{b_\alpha b_\alpha g^{(1)}}/g_s$, 
instead of the $\nu$-parameters. 

\begin{figure}[ht]
\includegraphics[width=8cm,clip]{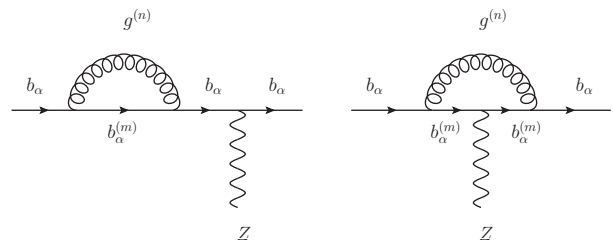}
\caption{The Feynman diagrams of 1-loop $Zbb$ vertex. 
}
\label{diagram}
\end{figure}

%
The Feynman diagrams of $Zbb$ vertex via the KK gluon exchange 
are shown in Fig.~\ref{diagram}. 
The vertex correction $\Delta g_\alpha^b~(\alpha=L,R)$ is given 
as follows: 
\begin{eqnarray}
\Delta g_\alpha^b &=& 
\frac{1}{\sqrt{4\sqrt{2}G_F m_Z^2}}
\left(
g^{bbZ}_\alpha \Sigma'(0)  - \Gamma_{b_\alpha}(m_Z^2)
\right), 
\label{dgbr}
\end{eqnarray}
where $\Sigma'(0)$ is the derivative of the self energy function 
of the external $b$-quark, whose mass is neglected. 
The scalar function $\Gamma_{b_\alpha}(m_Z^2)$ is the three point 
function of the $Z b_\alpha b_\alpha$ vertex at the momentum 
transfer $q^2 = m_Z^2$. 
The coupling of the $Z$-boson to $b_\alpha$ quarks is denoted by 
$g^{bbZ}_\alpha$. 
We note that the ultra violet divergences are cancelled between 
the self energy and vertex diagrams from each KK state. 
However, the 1-loop corrections become infinite when one takes 
the sum of the finite contributions from whole KK towers. 
We, therefore, need to introduce a cut-off scale $\Lambda$ to 
restrict the number of KK modes. 

%
The Na\"{\i}ve Dimensional Analysis
(NDA)\cite{Manohar:1984md,Scrucca:2003md}  
has 
been adopted to determine the cut-off scale $\Lambda$. 
In NDA, the cut-off scale $\Lambda$ is interpreted as an upper limit  
of energy scale in which a theory is perturbative. 
However, NDA tells us that 
the cut-off scale $\Lambda$ in the RS model does not much differ 
from the Planck scale, and the number of KK modes which is effective 
below $\Lambda$ is roughly $\sim 10^{15}$
\footnote{The cut-off scale $\Lambda$ in NDA for D-dimensional model is given by $\Lambda \sim \left( \left( 4\pi \right)^{D/2} \Gamma (D/2) /g_D^2 \right)^{1/(D-4)}$\cite{Scrucca:2003md}, where $g_D$ represents the D-dimensional gauge coupling. For the RS model, the cut-off scale is given by 
$\Lambda \sim l_5/g_5^2$ with $l_5=24\pi^3$ and the 5D gauge coupling is given by the 4D coupling as $g_5=g_4\sqrt{\pi r_c}$. Now we count the number of KK mode In the strong coupling limit of the 4D theory, i.e.,  $g_4^2 \sim 16\pi ^2$, 
If we then approximate the mass of the $n$-th KK mode $m_n \simeq n\pi k
\exp(-\pi kr_c)$, the number of KK modes below the cut-off scale could be $\Lambda / (\pi k\exp(-\pi kr_c) )\sim 10^{15}$.
}. 
Instead of NDA, therefore, we assume much lower cut-off scale 
(a few TeV) so that the finite number of KK modes is considered in the 
numerical analysis. 
\begin{figure}[ht]
\includegraphics[width=8.5cm]{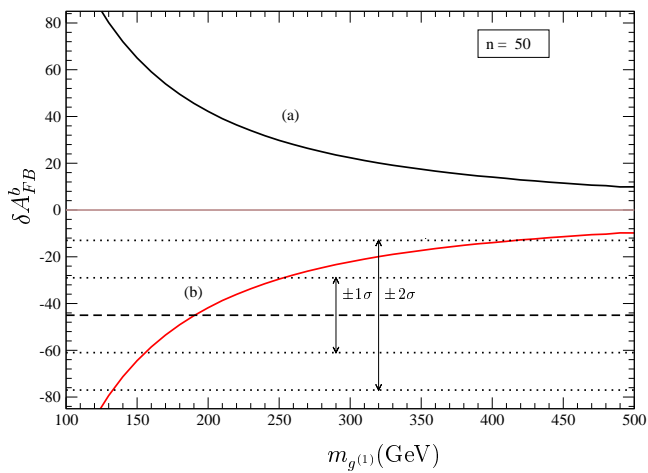}
\caption{
The KK gluon and the KK $b$-quark contributions to $\delta \afbb$ as a 
function of the 1st KK gluon mass, $\mgkk$. 
The upper and lower curves correspond to case (a) 
$(\xi_L,\xi_R)=(6, 0)$ and case (b) $(\xi_L,\xi_R)=(0,6)$, 
respectively.  
The horizontal dotted lines denote the allowed ranges of 
$\delta \afbb$ in 1- and 2-$\sigma$ level. 
}
\label{Amg}
\end{figure}

%
In Fig.~\ref{Amg} we show contributions from the KK gluons and the KK
$b$-quarks to $\delta \afbb$ as a function of the 1st KK gluon mass. 
The upper and lower curves correspond to case (a) 
$(\xi_L,\xi_R)=(6, 0)$ 
and case (b) $(\xi_L,\xi_R)=(0, 6)$, respectively. 
Note that only $\Delta g_L^b$ receives the KK gluon contribution 
in (a) while $\Delta g_R^b$ in (b). 
The results in the figure are obtained for the number of KK modes, 
$n=50$. 
The mass of the heaviest KK mode ($n=50$) depends on the mass of 
1st KK mode. 
For example, when $\mgkk = 200{\rm GeV}$, 
the mass of KK gluon and $b$-quarks for $n=50$ is $\sim 13{\rm TeV}$.  
The horizontal dotted lines denote the allowed range of $\delta \afbb$ 
in 1- and 2-$\sigma$ level as indicated in the figure. 
In the 1-loop correction to $\Delta g_\alpha^b$ (\ref{dgbr}), 
the sign difference comes from the $b_\alpha$-$b_\alpha$-$Z$ coupling 
$g_\alpha^b(\alpha=L,R)$. 
Since $g_\alpha^b \sim I_{3b} - Q_b \sin^2\theta_W$, we find 
the relative sign of $g_L^b$ and $g_R^b$ is opposite. 
This explains that the contribution to $\delta \afbb$ shows the opposite  
direction between case (a) and (b), 
since the coefficients of $\Delta g_L^b$ and $\Delta g_R^b$ have same
sign as shown in eq.(\ref{afbb}). 
Thus the KK gluon contribution to $\afbb$ is favored when the KK gluon 
couples dominantly to $b_R$. 
In the case of $(\xi_L,\xi_R)=(0, 6)$, the allowed range of the 1st KK 
gluon mass is $150 - 250{\rm GeV}$ in 1-$\sigma$ level 
($130 - 430{\rm GeV}$ in 2-$\sigma$ level). 
The range of KK gluon mass shifts when the couplings $(\xi_L,\xi_R)$ 
differ. 
A smaller value of $\xi_R$ lowers the favored range of KK gluon 
mass. For example, when $\xi_R=4$, the KK gluon mass 
which is allowed from $\afbb$ is $90 {\rm GeV}$-$150 {\rm GeV}$ 
in 1-$\sigma$. 

%
\begin{figure}[ht]
\includegraphics[width=8.5cm,angle=0,clip]{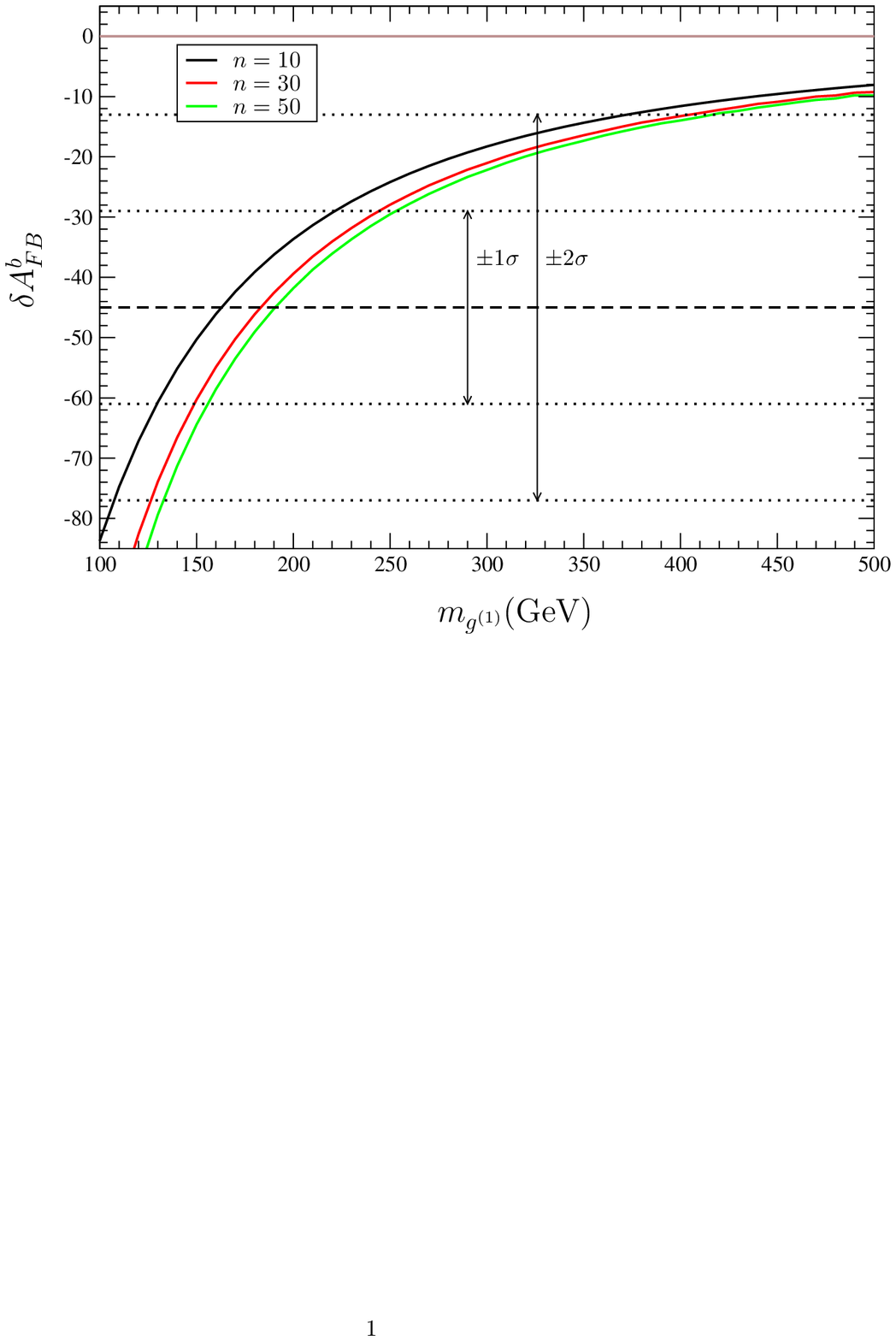}
\caption{
The KK gluon contributions to $\delta \afbb$ for the number of 
KK mode $n=10,30$ and 50 (from upper to lower curves). 
The couplings are $(\xi_L,\xi_R)=(0, 6)$. }
\label{n50}
\end{figure}
We have so far examined the KK gluon contribution to 
$\afbb$ for the number of KK mode $n=50$. 
The dependence of $\delta \afbb$ on the number of KK mode is shown 
in Fig.~\ref{n50} for $n=10,30$ and 50. 
The couplings are fixed at $(\xi_L,\xi_R)=(0, 6)$. 
We find that, when $m_{g^{(1)}}=200$GeV,  the difference of $\delta \afbb$
between $n = 10$ and $50$ is about few $10\%$ while it is few $\%$ between $n = 30$ and $50$.

\begin{table}[h]
\begin{tabular}{l|c|c|c}  \hline \hline 
 & Exp. & SM best fit & pull\\ \hline 
$R_b$ & $0.21629 \pm 0.00066$ & 0.21562 & 1.0 \\  \hline  
$A_b$ & $0.923 \pm 0.020$ & 0.935 & $-0.6$
\\ \hline \hline 
\end{tabular}
\caption{Experimental data and the SM best fit of $R_b$ and $A_b$ 
\cite{:2005ema}. 
The pull factor is defined as a deviation between data and the 
SM prediction normalized by the error. 
}
\label{tab}
\end{table}
The $Zbb$ vertex correction $\Delta g_R^b$ affects not only 
$\afbb$ but also other electroweak observables for $b$-quark jets 
-- for example, the partial decay rate $R_b$ and the left-right 
asymmetry $A_b$. 
Here let us briefly mention about correlations between 
$\Delta g_R^b$ and three observables $\afbb,R_b,A_b$. 
The experimental data and the SM prediction of $R_b$ and $A_b$ 
are summarized in Table.~\ref{tab}. 
As $\afbb$ (\ref{afbb}) , $R_b$ and $A_b$ can be expressed 
as~\cite{Hagiwara:1998yc}: 
\begin{eqnarray}
R_b({\rm NP}) 
&=& R_b({\rm SM}) - 0.78 \Delta g_L^b + 0.14 \Delta g_R^b, 
\\
A_b({\rm NP})  
&=& A_b({\rm SM}) - 0.30 \Delta g_L^b - 1.63 \Delta g_R^b. 
\end{eqnarray}
We consider $\Delta g_L^b=0 (\xi_L=0)$ in the following. 
When the shift of $\Delta g_R^b$ reduces the pull factor of 
$\afbb$ from $-2.8$ (SM best) to $-1.0$, we find that 
the pull factors of $(R_b,A_b)$ from 
their SM best fit $(1.0,-0.6)$ to $(-2.4,0.7)$. 
Then $\chi^2$ of three observables is reduced from 9.7 (SM best fit) 
to 7.5.  
From Fig.\ref{Amg}, the mass of 1st KK gluon which corresponds to 
the $-1.0\sigma$ of $\afbb$ data is about 250GeV for $\xi_R=6$. 
We conclude that, in a certain parameter space, the KK 
gluon contribution to the $Zbb$ vertex could explain 
the $\afbb$ data without affecting the current consistency 
of the other $b$-jet data, $R_b$ and $A_b$. 

To summarize, we have studied the KK gluon $g^{(n)}$ 
in the warped extra 
dimension model confronts the $\afbb$ data at the LEP experiments, 
which differs from the SM prediction about $-2.8\sigma$. 
We consider a scenario in which the coupling of $g^{(1)}$ and 
the zero-mode $b$-quark could be a few times larger than the QCD coupling 
depending on the localization position of the bulk wavefunction of 
$b$-quark. 
We examined the 1-loop correction of $Zbb$ vertex via the KK modes
exchange and found that the experimental data of $\afbb$ could be
explained when the KK gluon coupling to the right-handed $b$-quarks 
is sizable while to the left-handed $b$-quarks is highly suppressed. 
For example, the KK gluon with $\mgkk \sim 150-250{\rm GeV}$ is favored 
from the data when $\xi_L=0$ and $\xi_R=6$. 
We should mention that, however, since the parameter $\xi_R$ is defined
as a ratio of the coupling of KK gluon to $b_R$-quarks to the QCD 
coupling $g_s$, our choice $\xi_R=6$ is too large to be an expansion 
parameter of a perturbation theory. 
Therefore, our 1-loop calculations may be reliable when $\xi_R$ is  
much smaller ($\xi_R \ll 6$). 
In such a case, to explain the $\afbb$ data, the KK gluon mass 
is required to be sufficiently  
small, {\it e.g.}, $\mgkk \ll 100 {\rm GeV}$. 

A few comments are in order. 
In our scenario the KK gluon dominantly couples to $b_R$. 
Then, the production process of $g^{(1)}$ at hadron collider is 
$b\bar{b} \to g^{(1)}$. 
The production rate of $g^{(1)}$ is, therefore, suppressed 
even if $g^{(1)}$ is relatively light, $\sim O(100{\rm GeV})$. 
Note that the gluon fusion process $gg \to g^{(1)}$ is forbidden, 
because the zero-mode wavefunction in the fifth dimension is just 
a constant, and the 4D effective coupling of $g$-$g$-$g^{(1)}$ 
is zero 
due to the orthonormality condition of gluon wavefunctions. 
Since the decay of $g^{(1)}$ is possible only through 
$g^{(1)} \to b \bar{b}$, we compared the cross section 
$\sigma(p\bar{p} \to g^{(1)}+X) \times {\rm Br}(g^{(1)}\to b\bar{b})$ 
with the results given by CDF collaboration\cite{Aaltonen:2008dn}. 
When $(\xi_L,\xi_R)=(0, 6)$, constraint on $\mgkk$ from 
Tevatron is $\mgkk>157{\rm GeV}$ in 2-$\sigma$ level, which 
is consistent with the results obtained from $\afbb$ in this paper. 
The other possibilities of $g^{(1)}$ production at Tevatron are 
emission of $g^{(1)}$ from $b$ or $\bar{b}$ quark 
($p\bar{p}\to b\bar{b}g^{(1)}$), 
and a pair production of $g^{(1)}$ from gluon fusion, 
$gg \to g^{(1)} g^{(1)}$. 
After the decay of $g^{(1)}$, 
the final states are $b\bar{b}b\bar{b}$ in 
both cases and the excess of four $b$-jets event may be a signal at 
hadron collider experiments. 
Also the invariant mass distributions of two $b$-jets $m_{jj}$ may show
a peak at $m_{jj}=\mgkk$. 
Therefore, the analysis of four $b$-jet data at Tevatron is necessary 
to study further constraints on $g^{(1)}$.  
It is also interesting to study these processes at LHC, and the results
will be given in our forthcoming paper~\cite{ckmy2009}.

\begin{acknowledgments} 
The authors would like to thank Nobuhito Maru for reading manuscript and 
valuable comments. 
We also thank Kyoko Yoneyama for fruitful discussion. 
\end{acknowledgments}



\newpage
\begin{thebibliography}{99}
\bibitem{:2005ema}
LEP electroweak working group, 
  Phys.\ Rept.\  {\bf 427}, 257 (2006). 
%
\bibitem{Amsler:2008zzb}
  C.~Amsler {\it et al.}  [Particle Data Group],
  ``Review of particle physics,''
  Phys.\ Lett.\  B {\bf 667}, 1 (2008).

\bibitem{Peskin:1990zt}
  M.~E.~Peskin and T.~Takeuchi,
  Phys.\ Rev.\ Lett.\  {\bf 65}, 964 (1990); 
  Phys.\ Rev.\  D {\bf 46}, 381 (1992).
%
\bibitem{Hagiwara:1998yc}
  K.~Hagiwara,
  Ann.\ Rev.\ Nucl.\ Part.\ Sci.\  {\bf 48}, 463 (1998).
%
\bibitem{Altarelli:2001wx}
  G.~Altarelli, F.~Caravaglios, G.~F.~Giudice, P.~Gambino and 
				G.~Ridolfi,
  JHEP {\bf 0106}, 018 (2001). 
%
\bibitem{He:2002ha}
  X.~G.~He and G.~Valencia,
  Phys.\ Rev.\  D {\bf 66}, 013004 (2002)
  [Erratum-ibid.\  D {\bf 66}, 079901 (2002)]. 
%
\bibitem{Choudhury:2001hs}
  D.~Choudhury, T.~M.~P.~Tait and C.~E.~M.~Wagner,
  Phys.\ Rev.\  D {\bf 65}, 053002 (2002); 
  D.~Chang, W.~F.~Chang and E.~Ma,
  Phys.\ Rev.\  D {\bf 61}, 037301 (2000);
  F.~del Aguila, M.~Perez-Victoria and J.~Santiago,
  JHEP {\bf 0009}, 011 (2000). 
%
\bibitem{Randall:1999ee}
  L.~Randall and R.~Sundrum,
  Phys.\ Rev.\ Lett.\  {\bf 83}, 3370 (1999). 
%
\bibitem{Davoudiasl:1999tf}
  A.~Pomarol,
  Phys.\ Lett.\  B {\bf 486}, 153 (2000);

\bibitem{Davoudiasl:2000wi}
  H.~Davoudiasl, J.~L.~Hewett and T.~G.~Rizzo,
  Phys.\ Rev.\  D {\bf 63}, 075004 (2001). 

  C.~Csaki, J.~Erlich and J.~Terning,
  Phys.\ Rev.\  D {\bf 66}, 064021 (2002). 

\bibitem{Carena:2006bn}
  J.~L.~Hewett, F.~J.~Petriello and T.~G.~Rizzo,
  JHEP {\bf 0209}, 030 (2002);
  M.~S.~Carena, E.~Ponton, J.~Santiago and C.~E.~M.~Wagner,
  Nucl.\ Phys.\  B {\bf 759}, 202 (2006). 

\bibitem{Agashe:2003zs}
  K.~Agashe, A.~Delgado, M.~J.~May and R.~Sundrum,
  JHEP {\bf 0308}, 050 (2003). 


\bibitem{Djouadi:2006rk}
  A.~Djouadi, G.~Moreau and F.~Richard,
  Nucl.\ Phys.\  B {\bf 773}, 43 (2007)
  [arXiv:hep-ph/0610173].

\bibitem{Djouadi:2009nb}
  A.~Djouadi, G.~Moreau, F.~Richard and R.~K.~Singh,
  arXiv:0906.0604 [hep-ph].



\bibitem{Lillie:2007yh}
  B.~Lillie, L.~Randall and L.~T.~Wang,
  JHEP {\bf 0709}, 074 (2007). 

\bibitem{Grossman:1999ra}
  Y.~Grossman and M.~Neubert,
  Phys.\ Lett.\  B {\bf 474}, 361 (2000); 
  T.~Gherghetta and A.~Pomarol,
  Nucl.\ Phys.\  B {\bf 586}, 141 (2000). 
  
\bibitem{Manohar:1984md}
  A.~Manohar and H.~Georgi,
  Nucl.\ Phys.\  B {\bf 234}, 189 (1984).
  
\bibitem{Scrucca:2003md}
  C.~A.~Scrucca, M.~Serone and L.~Silvestrini,
  Nucl.\ Phys.\  B {\bf 669}, 128 (2003);
  B.~Grzadkowski and J.~Wudka,
  Phys.\ Rev.\  D {\bf 77}, 096004 (2008).
  

\bibitem{Aaltonen:2008dn}
  CDF Collaboration,
  http://www-cdf.fnal.gov/physics\\
  /new/qcd/bb\_SVT\_07/cdf8939\_bbjet\_highpt\_public.ps

\bibitem{ckmy2009}
  G.C.~Cho, Y.~Kanehata, N.~Maru and K. Yoneyama, in preparation. 
  
  


 \end{thebibliography}
\end{document}